\documentclass{article}

\usepackage{arxiv}

\usepackage[utf8]{inputenc} 
\usepackage[T1]{fontenc}    
\usepackage{hyperref}       
\usepackage{url}            
\usepackage{booktabs}       
\usepackage{amsfonts}       
\usepackage{nicefrac}       
\usepackage{microtype}      
\usepackage{lipsum}		
\usepackage{graphicx}
\usepackage{natbib}
\usepackage{doi}

\usepackage{amsmath,bm}

\usepackage{hyperref}
\usepackage{url}
\usepackage{placeins}

\usepackage{tikz}
\usetikzlibrary{calc}
\usetikzlibrary{positioning}

\title{Feature importance of socio-economic parameters in Tuberculosis modeling}
\date{}
\author{Andrei~Neverov \& Olga Krivorotko \\
Department of AI technologies for mathematical modeling
of\\ biological, socio-economic and environmental processes\\
Sobolev Institute of Mathematics SB RAS\\
Novosibirsk, Russia \\
\texttt{\{a.v.neverov,o.i.krivorotko\}@math.nsc.ru} \\
}

\renewcommand{\shorttitle}{Feature importance of soc.-econom. parameters in TB modeling}

\hypersetup{
pdftitle={Feature importance of socio-economic parameters in Tuberculosis modeling},
pdfsubject={q-bio.NC, q-bio.QM},
pdfauthor={Andrei~Neverov, Olga Krivorotko},
pdfkeywords={First keyword, Second keyword, More},
}

\begin{document}
\maketitle

\begin{abstract}
	This paper considers the problem of modeling epidemic outbreaks in different regions with a common model, that uses additional information about these regions to adjust its parameters and relieve us of mundanity of data collecting, and inverse problem solving for each region separately. To that end, we study tuberculosis and HIV dynamics in regions of Russian Federation from 2009 to 2023 in connection with number of socio-economic parameters. SIR-like model was taken and modified as a dynamic model for tuberculosis-HIV co-infection and inverse problem of transfer rates between compartments was solved, based on statistical data of diseases incidence. To shorten the list of socio-economic parameters we make use of Shapley vector that allows us to estimate importance of these parameters in reconstruction of differential model parameters using regression algorithms.
\end{abstract}


\section{Introduction}

One of the problems of the modeling of outbreak of endemic infectious is the lack of heterogeneity of the process in different regions~\citep{Aparicio2009}. This complication leads to heterogeneity of data and difficulties in development of a robust universal model with epidemiological inputs only. Previous works deal with this issue with introduction of reinfection~\citep{Feng2000}, households modeling~\citep{Aparicio2000} or socio-economic data~\citep{Azizah}, however they look into one local epidemic situation of a particular country, { 
    while these differences depend on location. It is not obvious what particular features of a location is connected to epidemic process, thus we implement feature importance method to estimate impact of these features on epidemic.
    }


{
    To describe epidemic dynamic SIR-like \citep{MathematicalModelingForecasting2020} or agent-based \citep{KrivorotkoSosnovskaiaKabanikhin2023} models are commonly used. The common idea of these approaches is to assign discrete number of states to the people in population and some rules or speeds of transition between states. 
    Definition of those states and transitions is characterized with set of epidemiological parameters (transmissibility, probability of progression to severe form, vaccine efficiency, etc.). Those parameters define whether stable solution with infected population exists, i.e. define whether or not some measures are to be done to eliminate the infection~\citep{Aparicio2009}.
}

{
    Our goal is to set and solve inverse problem of reconstruction of those epidemic parameters for Tuberculosis (TB) and HIV co-infection dynamics based on predetermined key socio-economic factors.
}

As socio-economic parameters are numerous and interconnected, we choose to work with part of it. There is evidence of major influence on epidemics dynamics from parameters depicting welfare of the population in the region~\citep{Duarte2021}. Deep study review~\citep{Andrade2018} indicates homogeneous effect of influence on such socio-economic parameters, as means of mitigating epidemics in middle and low-income regions. Close results were obtained for China~\citep{Li2015}. Based on several statistical analysis of tuberculosis infection in Russia~\citep{Podgayeva2011, Aminev2013}, we choose our set of socio-economic parameters.

For numerical estimation of connection of these parameters and epidemics { in terms of a regression model}, we present feature importance analysis based on Shapley values. { It allowed us to select the most important parameters. However, restoring epidemiological parameters from economical were accurate up to 10\% just in 10 regions out of 87.}

\section{Shapley values}

There are several basic feature importance approaches for machine learning algorithms. Linear models such as linear and logistic regression allow us to use their coefficients as feature importance. Tree based algorithms have variety approaches like Gini importance or number of times feature appears in the tree, however these methods are limited to tree based models and are empirical. The most straightforward method to measure feature importance is to use permutations of features, which may be computationally expensive and give misleading result when features are correlated, what is almost always the case in real world data. Other feature importance methods and model explainers use additional surrogate model from list of simple to interpret models, like linear models or tree based models, but these methods allow to estimate importance locally~\citep{molnar}.

Shapley values are a concept of solution of cooperative games with coalition, i.e. players may unite into coalitions and achieve better results in contrast to acting separately.

This concept answers the question of how to distribute payoffs of a game, as the players are making unequal contributions. First of all, this distribution of payoffs would be desirable to have the following properties:

\begin{enumerate}
    \item Linearity, i.e. a game may be split into several games and distribution of whole game equals sum of distribution of parts.
    \item Symmetry, i.e. it does not depend on players order.
    \item Efficiency, i.e. whole payoff is distributed without residual.
    \item Null player, i.e. player, that does not contribute to payoff, receives nothing.
\end{enumerate}

Counterintuitively, there exists only one such distribution of payoffs, the Shapley values~\citep{shapley:book1952}.

The Shapley value of a feature value is its contribution to the payout, weighted and summed over all possible feature value combinations

\begin{equation}
    s(i) = \sum\limits_{F \subset \{1,..,n\} \backslash j} \frac{|F|!(n-|F|-1)!}{n!}(p(F) - p(F \cup \{i\})
\end{equation}

where $s(i)\in\mathbf{R}$ is a Shapley value of an $i$ parameter, $F$ is a subset of parameter indexes, except for $i$, $n\in\mathbf{N}$ is a number of features, $p(F)\in\mathbf{R}$ is a payout of a game, i.e. prediction of a model, with a given set of features numbers $F$. In our case $p(F)$ is prediction of the number of infected by the machine learning algorithms, based on the set of input features $F$.

This method found a new field of application with advances in machine learning, where models are either simple and predictable, or cumbersome and uninterpretable.

The main idea is to treat parameters of the model as players whose goal is to optimize a target function. 

\FloatBarrier
\subsection{Data and preprocessing}

Parameters of the data obtained for every region are depicted in the following lists:

Epidemiological parameters are the number of:
\begin{enumerate}
\item TB cases;
\item HIV cases;
\item TB+HIV cases ;
\item TB infected HIV tested;
\item TB related deaths;
\item TB deaths in $<1$ year;
\item HIV cases in penitentiary organizations.
\end{enumerate}

\begin{figure}[h]
\begin{center}
\includegraphics[width=0.49\textwidth]{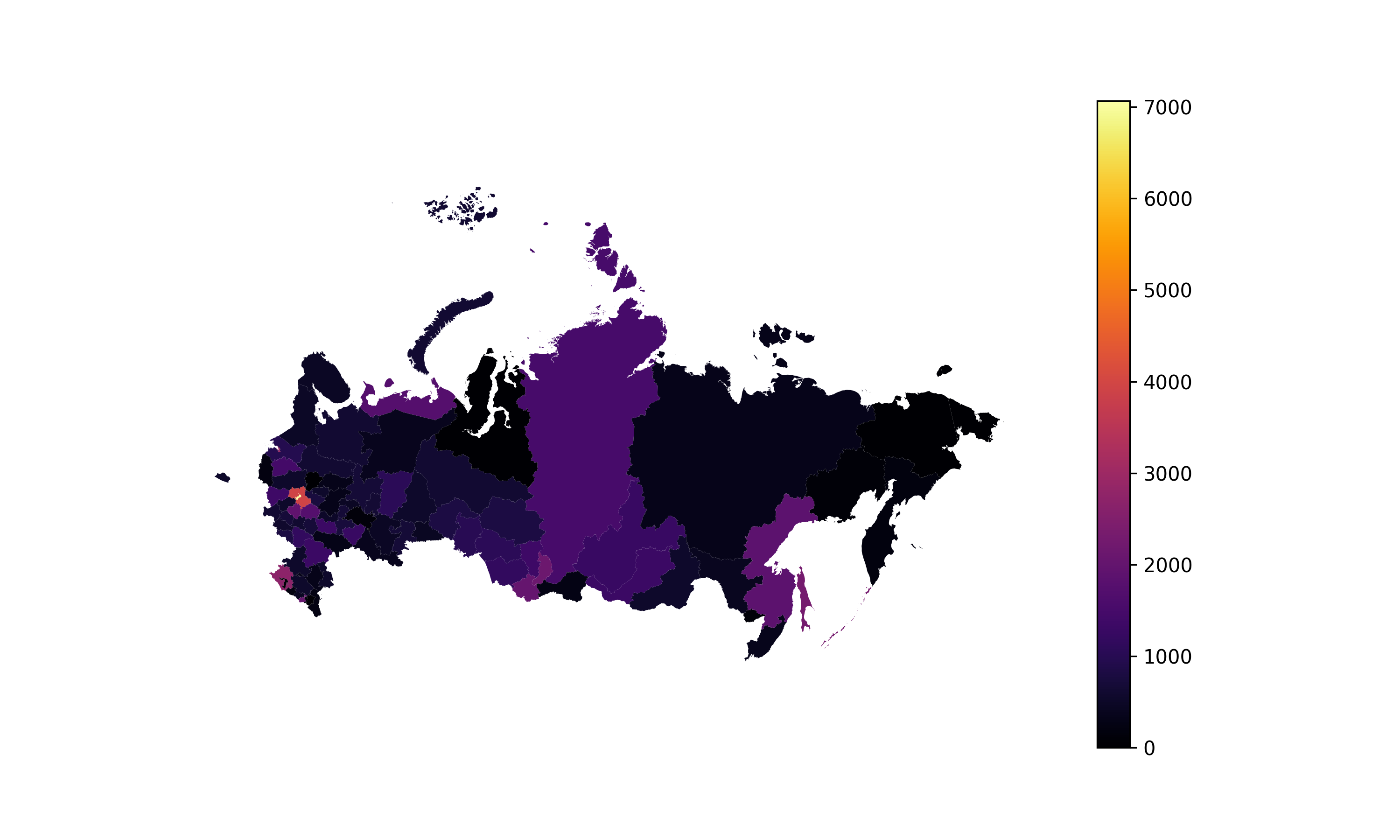}
\includegraphics[width=0.49\textwidth]{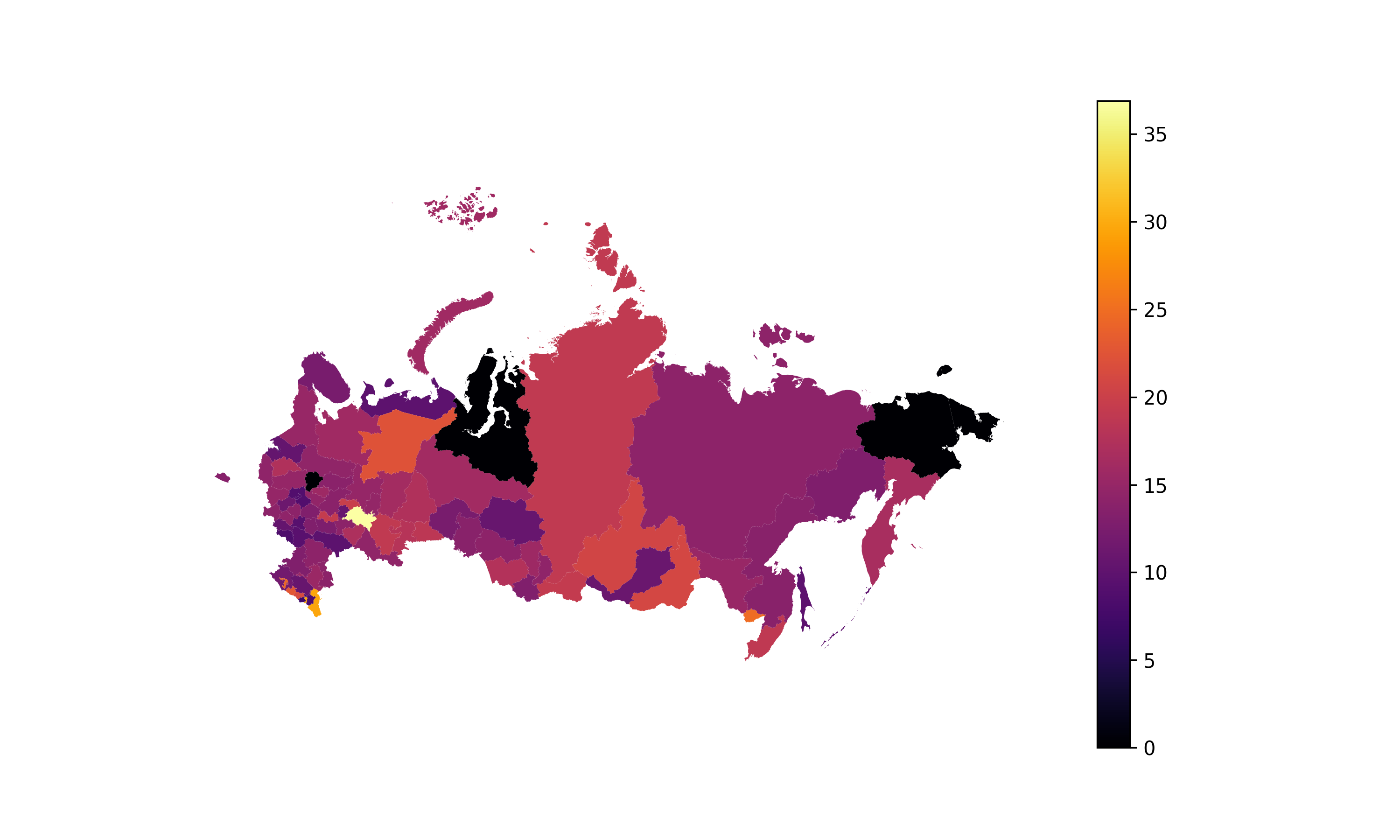}
\includegraphics[width=0.49\textwidth]{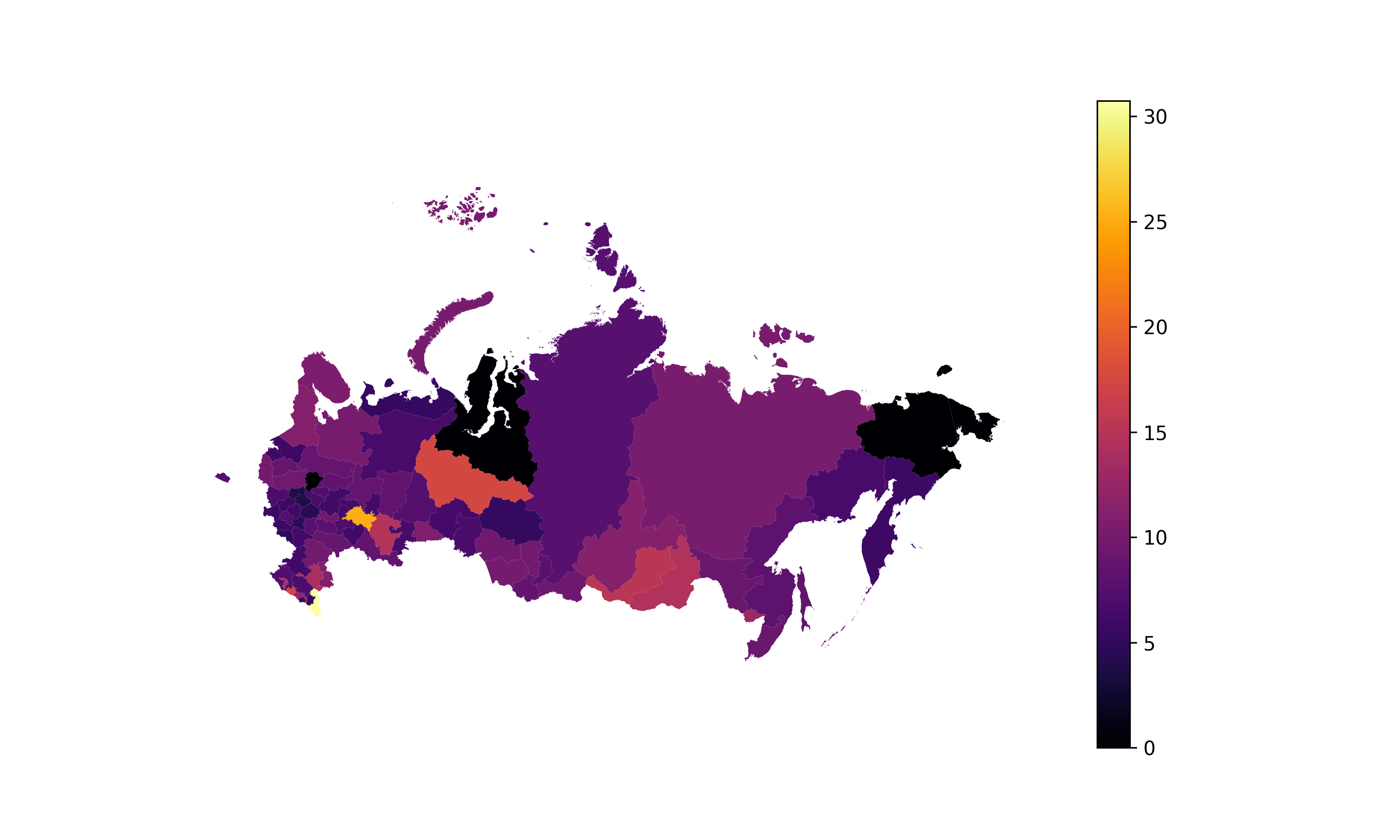}
\includegraphics[width=0.49\textwidth]{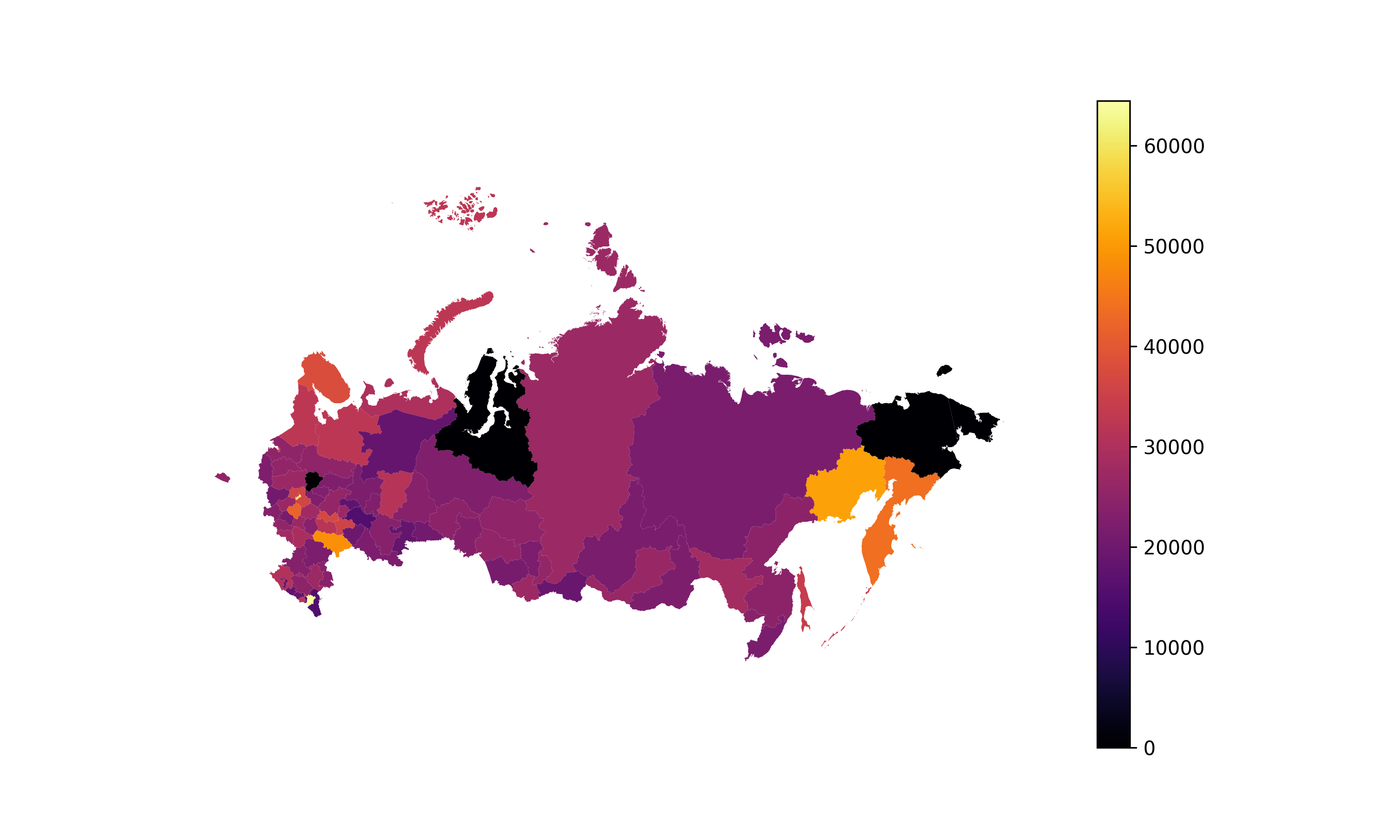}
\end{center}
\caption{The socio-economical statistics of Russian Federation regions. Top left - workforce, right - percent lower subsistence level, bottom left - unemployment, bottom right - mean income per capita.}
\label{fig:economical}
\end{figure}

Socio-economical parameters are (shown in Figure \ref{fig:economical}:
\begin{enumerate}
\item Population;
\item Unemployment;
\item Percent below subsistence level;
\item Median income per capita;
\item Size of workforce.
\end{enumerate}

All of the listed parameters are represented as time series since 2009 until 2019 with time-step of 1 year for each region of Russia.

\begin{figure}[h]
\begin{center}
\includegraphics[width=0.5\textwidth]{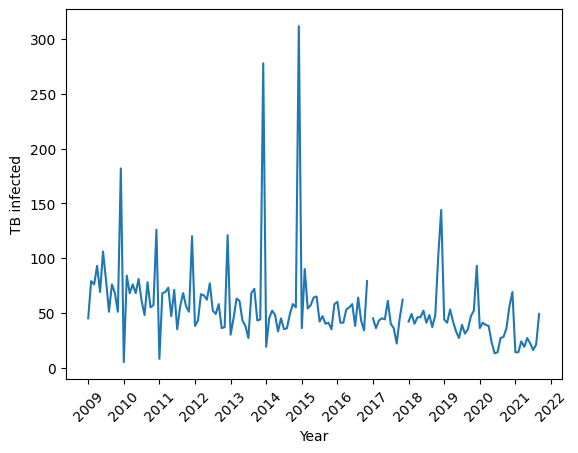}
\end{center}
\caption{Monthly new cases of tuberculosis infection in Tula region from 2009 to 2022.}
\label{fig:tula}
\end{figure}

When dealing with statistical data, we have to take into account the manner of its collection process. Effects of this are clearly seen in some particular parameters data, such as the number of new tuberculosis cases (See Fig. \ref{fig:tula}).

It is clear that seasonal outbreaks at the end of the years are not present in reality, because of the typical slow speed of tuberculosis infection. As we can not answer the question of how these outliers occur, we may not use such points of data. Moreover we may see, that sometimes data is missing, which coincides with cumulative parameters, thus we have to imitate data collection processes and skip some points of measurement, when solving inverse problem. Due to the lack of data, anomaly detection algorithms are not applicable in such scenario, and such skips are to be detected manually.

\section{ODE model and inverse problem}
 
As basis for our modeling we used the model proposed in \citep{Aparicio2009}, that includes AIDs, HIV, TB with latent and active stages and their co-infection. { This model is a SIR-like models, that start from works of~\citet{kermack}.} Addition of latent stage in TB infection is still debatable as it is still not clear how the latent TB progresses, what are the relevant average times of becoming latent and proceeding from latent to active TB. Moreover, the definition of latent TB has been changing throughout the last century, thus we have different viewpoints on what cases to call latent active or clear of TB \citep{Behr2021LatentTT, Cohen2019TheGP, Alexeeva_carrier, 10.1001/jama.2023.4899}. This puts us in a situation in which we do not know for sure if the definitions in different regions differ from each other.

In addition, latent TB group doubles the number of unknown parameters and adds unmeasured (or even immeasurable for current situation) group, that we need to restore. This situation generates situation where the compartment and coefficient of transfer from this compartment may not be reconstructed separately and uniquely, but only as some function of the pair (as their multiplication in most cases) \citep{Houben2016TheGB, Romanuha}. 
{
    For example, in \citep{mdpi}, estimation of latent group as 30\% of population as it is given by WHO, led to unrealistic peak in the beginning of the modeling and unrealistic mean times (around 200 years) of transition from latent to active TB. Thus we had to lower the fraction of latent infected to 1-5\%, depending on the region, what allowed us to restore admissible epidemiological parameters.
}

Everything mentioned above leads us to decision to simplify model and compile transfer through latent TB group to active TB group into one. We also remove treated compartment as it duplicates susceptible group. Consequently, we reintroduce treatment transfer from active TB to susceptible state. For the obtained model we have measurements of all compartments which allows us to reconstruct unknown parameters of transfer rates.


{
    To unite models with different epidemiological parameters, we restore these parameters from socio-economical parameters of the regions, with help of assumption that tuberculosis is social disease.
}

\subsection{ODE model}
The basis for describing the co-infection of TB and HIV was the SI model, characterized by a system of 5 differential equations: 
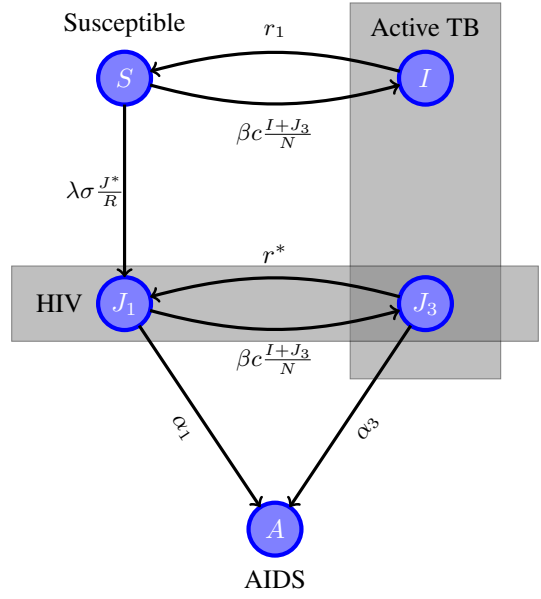
\begin{figure}[h]
\begin{center}
\begin{tikzpicture}
        [every node/.style={inner sep=0,outer sep=0}]
        \tikzset{node basic/.style={draw, ultra thick, blue, fill=blue!50, text=white, minimum size=2em}}
        \tikzset{node circle/.style={node basic, circle}}
        \tikzset{line basic/.style={very thick, ->}}
        


        \filldraw [fill=black!50, draw=black, semitransparent] (3, 1) rectangle (5, -4);

        \filldraw [fill=black!50, draw=black, semitransparent] (-1.5, -2.5) rectangle (5.5, -3.5);

        \node[node circle] (S) at (0, 0) {\(S\)};
        \node[node circle] (I) at (4, 0) {\(I\)};

        \node[node circle] (J1) at (0, -3) {\(J_1\)};
        \node[node circle] (J3) at (4, -3) {\(J_3\)};

        \node[node circle] (A) at (2, -6) {\(A\)};

        \node[above=0.2 of S] {Susceptible};
        \node[above=0.2 of I] {Active TB};
        \node[left=0.2 of J1] {HIV};

        \node[below=0.2 of A] {AIDS};

        \draw[line basic] (S) to [bend right=15] node [below=0.2, sloped] {{\small $\beta c \frac{I+J_3}{N}$}} (I);

        

        \draw[line basic] (S) to node [left] {\small $\lambda \sigma \frac{J^*}{R}$} (J1);
        

        
        \draw[line basic] (J1) to [bend right=15] node [below=0.2, sloped] {\small $\beta c \frac{I+J_3}{N}$} (J3);

        \draw[line basic] (I) to [bend right=15] node [above=0.2, sloped] {\small $r_1$} (S);
        \draw[line basic] (J3) to [bend right=15] node [above=0.2, sloped] {$r^*$} (J1);

        \draw[line basic] (J1) to node [below=0.2, sloped] {\small $\alpha_1$} (A);
        \draw[line basic] (J3) to node [below=0.2, sloped] {\small $\alpha_3$} (A);

\end{tikzpicture}
\end{center}
\caption{Flu diagram of ODE model of TB and HIV co-infection.}
\label{fig:model-structure}
\end{figure}


\begin{eqnarray}\label{tb-hiv-model-eq}
    \left\{\begin{array}{ll}
        \dfrac{dS}{dt} & = \Lambda - \beta c S \dfrac{I+J_3}{N} - \lambda \sigma S \frac{J^*}{R} - \mu S + r_1 I; \\[6pt]
        \dfrac{dI}{dt} & = \beta c S \dfrac{I+J_3}{N} - (\mu+d+r_1)I; \\[6pt]
        \dfrac{dJ_1}{dt} & = \lambda \sigma S \dfrac{J^*}{R} - \beta c J_1 \dfrac{I+J_3}{N} - 
    (\alpha_1 +\mu)J_1; \\[6pt] 
        \dfrac{dJ_3}{dt} & = \beta c J_3 \dfrac{I+J_3}{N} - (\alpha_3 +\mu+d_2+r^*)J_3; \\[6pt]
        \dfrac{dA}{dt} & =\alpha_1 J_1 + \alpha_3 J_3 - (\mu + f)A,
    \end{array}\right.
\end{eqnarray}

with initial conditions
\begin{eqnarray}\label{init_cond}
    S(0)=S_0, \; I(0)=I_0,\; J_1(0)=J_{1_0},\; J_3(0)=J_{3_0},\; A(0)=A_0.
\end{eqnarray}

Here $N = S(t)+I(t)+J_1(t)+J_3(t)+A(t)$ is the size of the population under study, $R = N - I(t) - J_3(t) - A(t) = S(t)+J_1(t)$~-- population without active phase of tuberculosis and AIDS, $J^* = J_1(t)+J_3(t)$~-- HIV-infected people who have not yet developed the AIDS stage. The population is divided into the following groups:
\begin{itemize}
\itemsep0em
 \item $S$~-- susceptible non-immunized population,
 \item $I$~-- patients with active form of tuberculosis,
 \item $J_1$~-- HIV infected individuals,
 \item $J_3$~-- infected with HIV and active form of tuberculosis,
 \item $A$~-- patients with AIDS.
\end{itemize}

Within the model, the entire population, regardless of status, is subject to natural mortality with the parameter $\mu$. A susceptible individual from group $S$ passes to the stage of active TB $I$ after contact with patient $I$ or $J_3$ at a rate of $\beta c \frac{I+J_3}{N}$. Parameter $\beta c$ is the probability $\beta$ of contracting tuberculosis upon contact with a person with an active form of tuberculosis { multiplied by contact rate $c$, that we assume to depend on socio-economical parameters of the region}. The $\frac{I+J_3}{N}$ multiplier reflects the proportion of the population that is infectious with tuberculosis to a susceptible population. A susceptible individual $S$ can transition to the HIV-infected stage from state $J_1$ after contact with HIV-infected groups $J^*=J_1+J_3$ at a rate of $\lambda \sigma \frac{J^*}{R} $. The parameter $\lambda \sigma$ describes the probability of contracting HIV infection when in contact with a person with HIV. Individuals infected from cells $I$, $J_3$ and $A$ are considered too ill to remain sexually active and therefore cannot transmit HIV through sexual intercourse. Thus, $R = N - I - J_3 - A$ represents the active population susceptible to infection, and the multiplier $\frac{J^*}{R}$ reflects the proportion of the population that is infectious with HIV for those who are susceptible.
{ 
    The similar description with table of parameters with units and typical values may be found in \citet{mdpi}.}

Patients with active tuberculosis $I$ are cured and go to susceptible state $I$ at a rate of $r_1$, or die with probability $\mu$ in the case of natural death and with probability $d$ in the case of death from tuberculosis.

As has already been found out, $J_1$ are included in the HIV-infected group from the group of susceptible $S$ at a rate of $\lambda \sigma \frac{J^*}{R}$. People infected with HIV can move to the stage of HIV-infected and active TB patients $J_3$ at a rate of $\beta c \frac{I+J_3}{N}$ if they have come into contact with people infected with tuberculosis from groups $I$ and $J_3$. An HIV-infected person can enter the AIDS stage $A$ with a probability of $\alpha_1$ or die with a probability of $\mu$.

Patients with active tuberculosis and infected with HIV $J_3$ can recover and go to the state of $J_2$ at a rate of $r^*$, or die with a probability of $\mu$ in the case of natural death and with a probability of $d^*$ in the case of death from tuberculosis. They also enter the AIDS stage at the rate of $\alpha_3$.

\FloatBarrier
\subsection{Inverse problem}\label{sec:inv}

To construct epidemiological model, based on socio-economical parameters, we start with description of epidemiological data through lens of socio-economical parameters with machine-learning tools, to eliminate the most unrelated and non-informative parameters.

We present comparison of set of basic machine learning models, that were tested on data from 2011 to 2019 years, with 50/50 split for train and test data. Resulting errors were averaged across all the regions and are presented in Table \ref{models-table}.

\begin{table}[h]
\caption{Tested statistical models}
\label{models-table}
\begin{center}
\begin{tabular}{ll}
\multicolumn{1}{c}{\bf MODEL}  &\multicolumn{1}{c}{\bf MEAN RELATIVE ERROR}
\\ \hline \\
Linear regression         & 0.102 \\
Ridge regression & 0.109\\
Gaussian process with dot product and white noise kernels & 0.030\\
Gaussian process with RBF kernel& 0.101\\
Tree& 0.135
\end{tabular}
\end{center}
\end{table}

\begin{figure}[h]
\begin{center}
\includegraphics[width=0.5\textwidth]{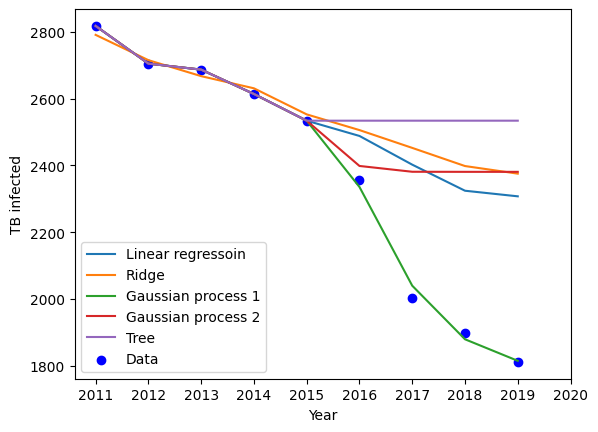}
\end{center}
\caption{Representative results for prediction of models through the example of Krasnoyarsk region data.}
\label{fig:predict}
\end{figure}

Based on Figure \ref{fig:predict} that shows representative example of resulting prediction for listed models, we choose Gussian process with White noise and Dot product kernels approximation to be our base model for evaluating Shapley values.

Then we solve the inverse problem of reconstruction unknown parameters $q = \{\beta c, \lambda \sigma, r_1, r^*, k\}$ with minimisation of a misfit functional:

\begin{equation}\label{misfit2}
    J(\vec q) = \sum_{t_i=2007}^{2020} \frac{|I^d(t_i) - I^m(t_i, \vec q)|^2}{M_I^2} + \frac{|J_1^d(t_i) - J_1^m(t_i, \vec q)|^2}{M_{J_1}^2} + \frac{|J_3^d(t_i) - J_3^m(t_i, \vec q)|^2}{M_{J_3}^2}. 
\end{equation}

Here $\{I^d, J_1^d, J_3^d\}$ are statistical data and $\{I^m, J_1^m, J_3^m\}$ is a simulation result of the model respectively, $M_s = \max\limits_{{t_i}}\{s^d(t_i)\}$ for the index $s \in \{I, J_1, J_3\}$ is a normalizing term.

We apply Tree Parsen Estimator as global optimization algorithm based on Bayesian approach, implemented in Optuna Python package. 


\FloatBarrier
\section{Numerical experiments}

{
    The numerical experiments were conducted the following way:

    \begin{enumerate}
        \item For each region solve inverse problem for the simplified TB-HIV problem~(\ref{tb-hiv-model-eq})-(\ref{init_cond}) with minimization of functional \ref{misfit2} by Bayesian-type approach~\citet{mdpi}, obtaining epidemiological parameters including $\beta$.
        \item Solve regression problem for $\beta$ with socio-economical parameters for collection of all regions data (we use Gaussian process for regression).
        \item Evaluate Shapley values based on results of regression models.
        \item Pick the parameters with the largest influence, based on dispersion of Shapley values.
    \end{enumerate}
    }

\begin{figure}[h]
\begin{center}
\includegraphics[width=1\textwidth]{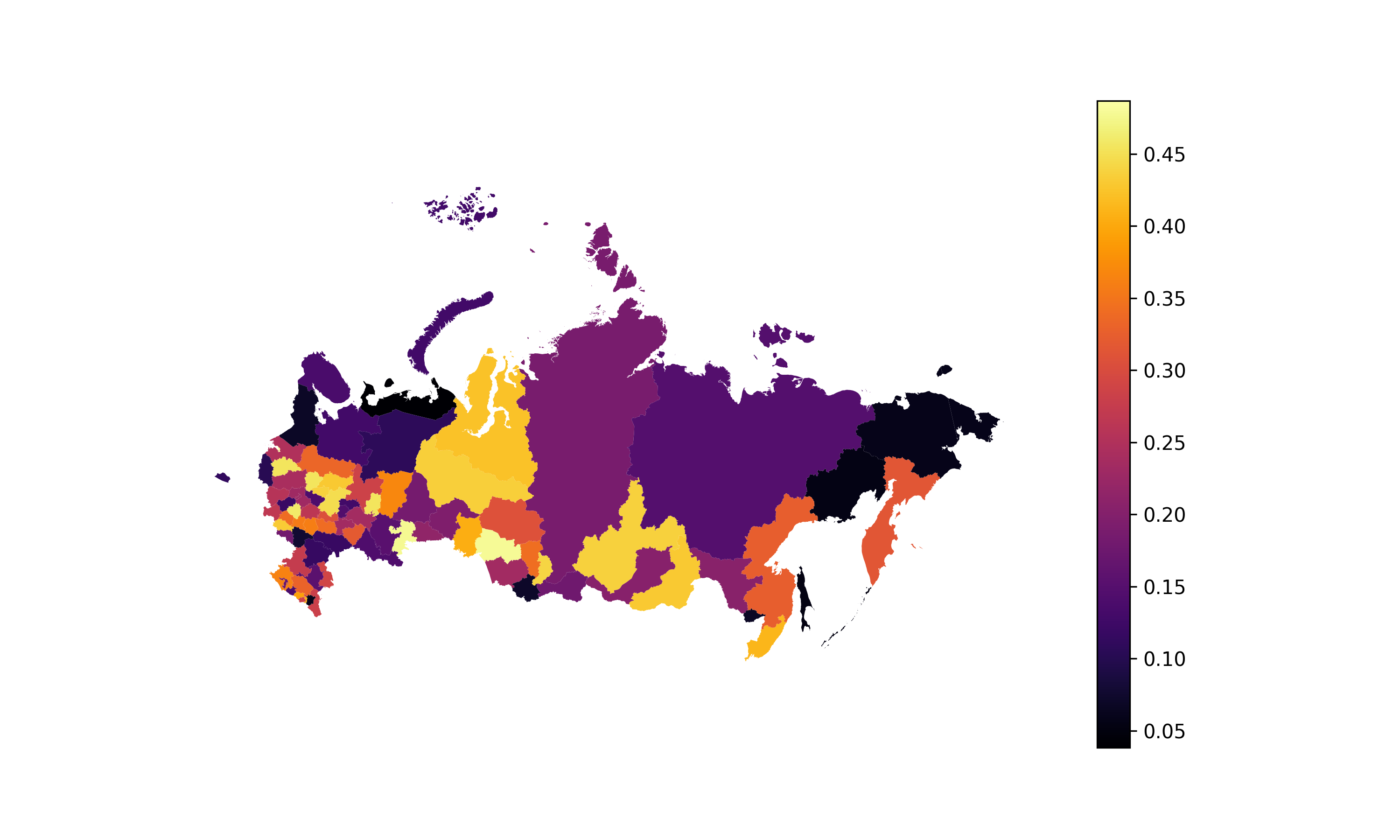}
\end{center}
\caption{Reconstructed $\beta$ for every region.}
\label{fig:map}
\end{figure}

Reconstructed contagiousness parameters $\beta$ for every region with the algorithm presented in Section \ref{sec:inv} are shown in figure \ref{fig:map}.

Shapley values for chosen Gaussian process model, based on socio-epidemiological parameters were obtained. For all the following figures, red color of dot represents above average value of feature, and in contrast blue represents the lower values.

\begin{figure}[h]
\begin{center}
\includegraphics[width=1\textwidth]{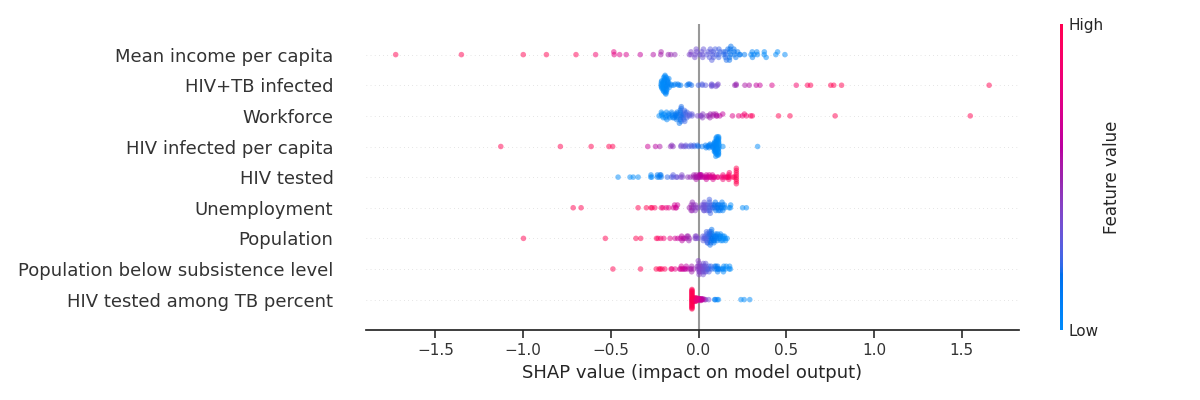}
\end{center}
\caption{Shapley values of socio-economic parameters in relation to $\beta c$.}
\label{fig:shap}
\end{figure}

From distribution of the points we may deduce analogue of correlation between parameters and infection rate $\beta c$. The model showed positive correlation of infection rate with number of detected HIV+TB infected, volume of workforce and number of HIV tested, and negative correlation with mean income, number of HIV infected per capita, unemployment, total population and population below subsistence level.

These results are unexpected for mean income and population below subsistence level as TB is more common to find in more unfavourable areas. However, these results do not show the causation, but causation from the point of view of chosen base regression model.

Levels of impact obtained in Figure \ref{fig:shap} let us choose the most impactful parameters to recover $\beta c$, thus we choose mean income per capita, workforce, number of HIV tested unemployment and total population for reconstruction of beta.

However, the regression was able to describe satisfyingly just 10 out of 87 parameters beta, thus making the results above applicable to only this small portion of regions. These regions are:

Kamchactka krai, Krasnoyarsk krai, Leningrad oblast, Republic of Dagestan, Republic of Mordovia, Republic of Northern Osetia, Rosvov oblast, Samara oblast, Smolensk oblast, Tver oblast, Tomsk oblast.

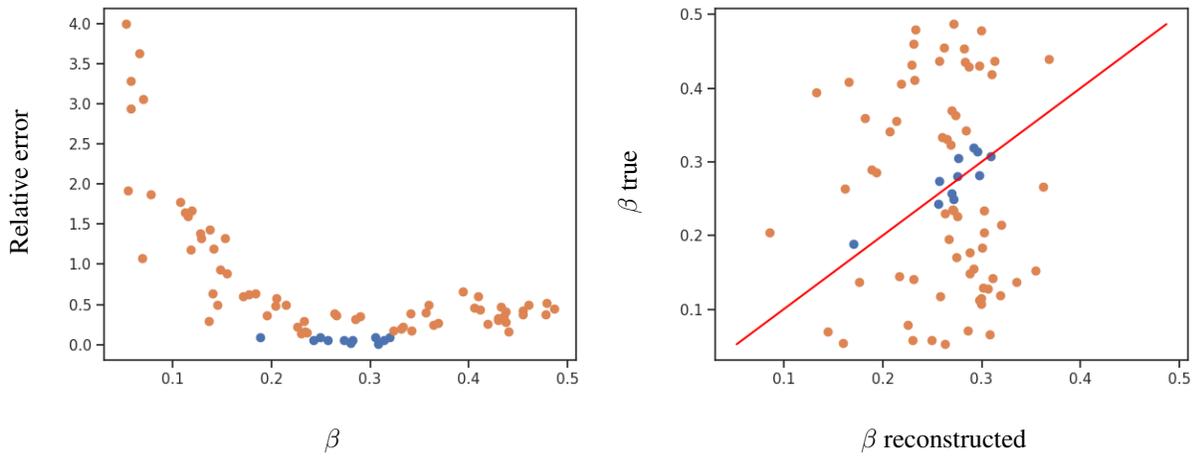
\begin{figure}[!h]
\begin{minipage}{0.49\textwidth}
\begin{tikzpicture}
  \node (img)  {\includegraphics[width=\textwidth]{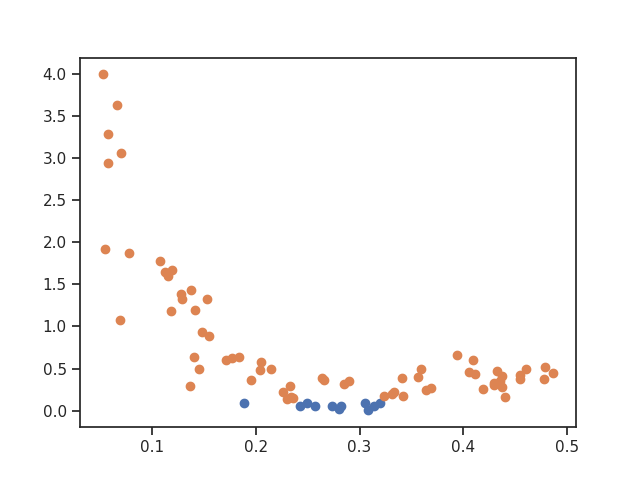}};
  \node[below=of img, node distance=0cm, yshift=1cm] {$\beta$};
  \node[left=of img, node distance=0cm, rotate=90, anchor=center,yshift=-1cm] {Relative error};
 \end{tikzpicture}
\end{minipage}%
\begin{minipage}{0.49\textwidth}
\begin{tikzpicture}
  \node (img)  {\includegraphics[width=\textwidth]{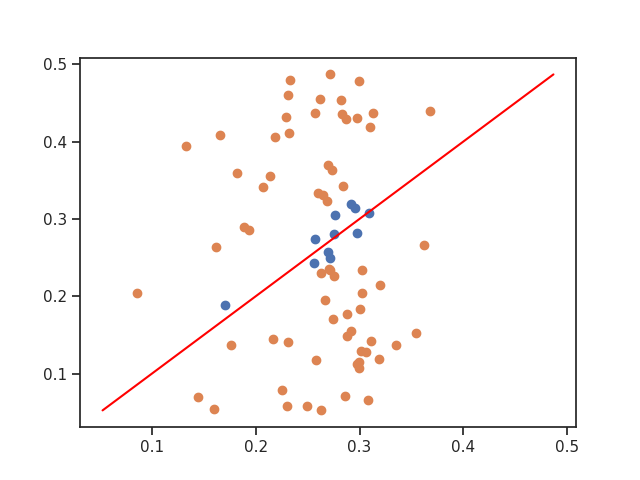}};
  \node[below=of img, node distance=0cm, yshift=1cm,] {$\beta$ reconstructed};
  \node[left=of img, node distance=0cm, rotate=90, anchor=center,yshift=-1cm] {$\beta$ true};
\end{tikzpicture}
\end{minipage}%
\caption{Results of reconstruction of $\beta$ from socio-economical parameters. Resulting relative error for each region (left) and comparison plot of true $\beta$ versus $\beta$ reconstructed (right). Blue dots show regions with error less than 10\%, orange dots are the rest. Red line represents area where $\beta$ true and reconstructed are the same.}
\label{fig:beta_res}
\end{figure}

The complete distribution of betas are presented in Fig.\ref{fig:beta_res}. In it we may notice that regression model worked much better for median $\beta$. The worst relative error is for the smallest $\beta$ (right Fig. \ref{fig:beta_res}) what may be easily explained with the same absolute level (see distance from points to red line in left Fig. \ref{fig:beta_res}).

\section{Conclusion}

The differential model of tuberculosis-HIV co-infection dynamic has been modified in order to obtain stable and unique solution of the inverse problem by eliminating group of latent infected. The coefficient inverse problem has been solved for 87 regions of Russian Federation. Based on Shapley values analysis, we have picked the most influential socio-economical parameters for regression of infection rate $\beta$. However, the results of regression have been satisfactory only for 10 regions, thus the model is to be further refined to have broader application in terms of regions and forecast.

\bibliographystyle{unsrtnat}
\bibliography{references}

\end{document}